# Layer-by-Layer Epitaxy of Multilayer MoS$_2$ Wafers

Qinqin Wang[1,2,†], Jian Tang[1,2,†], Xiaomei Li[1,2], Jinpeng Tian[1,2], Jing Liang[4], Na Li[1,3], Depeng Ji[3], Lede Xian[3], Yutuo Guo[1,2], Lu Li[1,2], Qinghua Zhang[1,2], Yanbang Chu[1,2], Zheng Wei[1,2], Yanchong Zhao[1,2], Luojun Du[1], Hua Yu[1,3], Xuedong Bai[1,2], Lin Gu[1,2], Kaihui Liu[4], Wei Yang[1,2], Rong Yang[1,2,3], Dongxia Shi[1,2] & Guangyu Zhang[1,2,3]*

[1] Beijing National Laboratory for Condensed Matter Physics and Institute of Physics, Chinese Academy of Sciences, Beijing 100190, China
[2] School of Physical Sciences, University of Chinese Academy of Sciences, Beijing 100190, China
[3] Songshan Lake Materials Laboratory, Dongguan 523808, China
[4] Collaborative Innovation Center of Quantum Matter and School of Physics, Peking University, Beijing 100871, China

[†] Authors contribute equally to this work.
* Corresponding author. Email: gyzhang@iphy.ac.cn

**Two-dimensional (2D) semiconductor of MoS$_2$ has great potential for advanced electronics technologies beyond silicon[1-9]. So far, high-quality monolayer MoS$_2$ wafers[10-12] are already available and various demonstrations from individual transistors to integrated circuits have also been shown[13-15]. In addition to the monolayer, multilayers have narrower band gaps but improved carrier mobilities and current capacities over the monolayer[5,16-18]. However, achieving high-quality multilayer MoS$_2$ wafers remains a challenge. Here we report the growth of high quality multilayer MoS$_2$ 4-inch wafers via the layer-by-layer epitaxy process. The epitaxy leads to well-defined stacking orders between adjacent epitaxial layers and offers a delicate control of layer numbers up to 6. Systematic evaluations on the atomic structures and electronic properties were carried out for achieved wafers with different layer numbers. Significant improvements on device performances were found in thicker-layer field effect transistors (FETs), as expected. For example, the average field-effect mobility ($\mu_{FE}$) at room temperature (RT) can increase from ~80 cm$^2\cdot$V$^{-1}\cdot$s$^{-1}$ for monolayer to ~110/145 cm$^2\cdot$V$^{-1}\cdot$s$^{-1}$ for bilayer/trilayer devices. The highest RT $\mu_{FE}$=234.7 cm$^2\cdot$V$^{-1}\cdot$s$^{-1}$ and a record-high on-current densities of 1.704 mA·µm$^{-1}$ at $V_{ds}$=2 V were also achieved in trilayer MoS$_2$ FETs with a high on/off ratio exceeding 10$^7$. Our work hence moves a step closer to practical applications of 2D MoS$_2$ in electronics.**

Since the successful exfoliation of 2D MoS$_2$[19], these ultrathin semiconductors have attracted great attentions in the field of electronics[4,10,13-15,20]. Tremendous efforts have been devoted to exploring their scaled-up potentials, including both wafer-scale synthesis of high-quality materials and application of them in large-area devices, with a specific focus on the monolayer MoS$_2$ (ML-MoS$_2$)[10,11,20-23]. Up to now, high-quality ML-MoS$_2$ wafers are already available from various growth approaches including chemical vapor deposition (CVD)[11,12] and metal organic CVD (MOCVD)[10]. Depending on the growth approaches and substrates, the MOCVD/CVD ML-MoS$_2$ films are generally stitched from random/aligned domains with sizes featured at 1/100 micron level and have a state-of-the-art room temperature electron mobility of ~30/~70 cm$^2 \cdot$V$^{-1} \cdot$s$^{-1}$ in average, an electronic quality comparable with or even better than the exfoliated monolayers.

In terms of a further improvement of the electronic quality of the large-scale 2D-MoS$_2$, structural imperfections should be eliminated as much as possible; however, there is not much space left for monolayer MoS$_2$ after ten years of synthesis optimizations in this field. Another direction is to switch to multilayer MoS$_2$, e.g. bilayers or trilayers, since they have intrinsically higher electronic quality than monolayers[16,18,24-27]. Indeed, with increased number of MoS$_2$ layers, decreased band gaps but enhanced electron mobilities and current densities have been demonstrated in exfoliated or CVD flakes[16,27,28]. However, it currently remains a significant challenge to produce high-quality and large-scale MoS$_2$ multilayers with well-controlled number of layers. Previously, CVD and sulfurization have been used to produce multilayer MoS$_2$ in form of flakes. While those flakes are of good crystal quality, their sizes are small, typically less than ~300 μm[25,29]. Large-scale multilayer MoS$_2$ films have been also synthesized, e.g. from sulfurization of precoated Mo/MoO$_3$ films[30] and atomic layer deposition (ALD)[31]. As-produced films are typically polycrystalline with the many randomly oriented domains in sizes of less than 100 nm and include co-existence of different layer thicknesses. Such poor crystalline quality, subjected to bad domain stitching and less control on the number of layers, leads to low electronic performances even worse than those achieved in MoS$_2$ monolayers[32-35].

Generally, to produce MoS$_2$ multilayers, the best practice is to begin with monolayers, then increase their thicknesses by gradually growing additional layers. However, considering the case of free-standing MoS$_2$, this route is problematic in the thermodynamic point of view. The surface energy of free-standing MoS$_2$ increases with the number of layers[36,37], it is thus energetically unfavorable to increase additional layers[38]. This fundamental thermodynamic limitation is likely to prevent large area multilayer MoS$_2$ with well-controlled layer numbers from being demonstrated previously. It is expected that this thermodynamic limitation might be overcome by engineering the surface energy of MoS$_2$ via the proximity effect.

In this work, we developed a new technique, i.e. layer-by-layer epitaxy, to grow high-quality 4-inch multilayer MoS$_2$ wafers with controlled number of layers. By using sapphire (0001) as the starting substrate, we successfully achieved the growth of uniform NL-MoS$_2$ (N=1, 2, 3), where N is the number of layers, in a layer-by-layer manner. All sapphire wafers used in our growth are 4-inch wafers cut along the zero-degree plane, or C-Cut, which were vacuum annealed at ~1000 ºC to form atomically flat surfaces before epitaxy. Note that sapphire wafers are cheap and widely used for various semiconductor thin film epitaxy; and sapphire (0001) surface is so far one of the best substrate for MoS$_2$ epitaxy due to a negligible lattice mismatch. Obviously, two processes are involved in this layer-by-layer growth, i.e. heteroepitaxy of 1st layer on sapphire and homoepitaxy of (N+1)th layer on NL with N > 0, as illustrated in Figure 1a.

Both heteroepitaxy and homoepitaxy growth were performed in a multi-source oxygen-enhanced CVD system. As shown in our previous studies, this new CVD approach for monolayer $MoS_2$ growth features greatly enhanced growth rate and excellent film uniformity across the entire 4-inch sapphire surface (benefited from the stable and uniform S- and Mo- source supply during the growth process)[11]. Usually, the 1st layer epitaxy on sapphire starts from nucleation at multiple sites, proceeds with the edge growth of those nuclei and eventually reaches a layer completion (i.e. full coverage on the substrate surface) via the domain-domain coalescence mechanism.

Note that monolayer $MoS_2$ growth on sapphire or $SiO_2$ substrates follows a unique self-limiting process[21] in which additional layers can hardly be nucleated on the monolayer till its completion. After completing the 1st layer, we then increased both temperature of Mo-source ($T_{Mo}$) and substrate ($T_{substrate}$) to enable dense nucleation of the 2nd layer (see Methods). A higher $T_{Mo}$, in other words, a higher Mo-source flux, is found to be beneficial to achieve a higher nucleation density of the 2nd layer.

As shown above, the dedicate control of the growth kinetic process, e.g., nucleation and edge growth, is the key to achieve continuous layer epitaxy. In our growth tests, we achieved $MoS_2$ wafers with N up to 6. It was noticed that the ideal 2D growth mode is difficult to keep for $N^{th}$ layer when $N \geq 3$, leading to appearance of additional mono- or multilayer domains on NL-/$MoS_2$. Such failure is more and more significant with increasing N and the growth mode evolves gradually from 2D to 3D, in consistence with the classical Stranski-Kranstanov growth mode[40]. This layer-dependent growth mode evolution could be attributed to several reasons. Firstly, surface proximity effect reduces quickly for those thicker layers with upper surfaces farther away the sapphire surface. Besides, once the additional layers appear, their presence would amplify in the growth of subsequent layers.

Since as-grown $MoS_2$ films are very uniform for mono-/bilayers and quite uniform for trilayers across entire 4-inch wafers, we thus mainly focus on the bilayer and trilayer samples in the following characterizations. Fig. 1b shows typical optical images of the as-grown 4-inch mono-, bi- and trilayer $MoS_2$ wafers. Fig. 1c-h show typical zoom-in optic and atomic force microscope (AFM) images from these wafers, indicating the full coverage and very clean surfaces. The trilayer continuous films have certain additional small quadrilayer domains, and the coverage of them is ~30 %. The layer numbers were further confirmed by high resolution cross-section high-angle annular dark-field scanning transmission electron microscopy (HAADF-STEM) imaging (Fig. 1i-k). We can see clearly that each layer consists of one layer Mo and two layer S atoms with layer thickness of ~0.62 nm and the interface between the adjacent layers is atomically clean and sharp, reflecting the superiority of epitaxy.

In order to elucidate the layer stacking orders in these multilayer $MoS_2$ wafers, we further performed atomic structure characterizations by STEM. As shown in Fig. 2, there are two stacking orders in our bilayer samples, i.e. AA stacking (2L-AA, 3R phase) and AB stacking (2L-AB, 2H phase) and the corresponding atomic configurations are shown in Fig. 2a. Fig. 2b/2c show STEM images of a typical AA/AB-stacked bilayer $MoS_2$. Note that the AA stacked layers have no inversion symmetry while the AB stacked layers have. Fig. 2d&e also show the TEM images of a bilayer $MoS_2$ film with a grain boundary. AA and AB stacked domains can be clearly distinguished and these two different stacking domains can coalesce together without any disconnect gap, revealing a crystalline continuity. Fig. 2f shows the selected-area electron diffraction (SAED) pattern at the grain boundary area, exhibiting only one set of hexagonal diffraction spots, as expected. We also characterized the trilayer samples. Different from the bilayer case, the stacking orders in trilayers are much more complicated. AAA, AAB/ABB, and ABA stacking configurations all exist, as show in Fig. 2 g-i. All these STEM images for bi- or trilayers reveal our epitaxial multilayer films having excellent lattice alignments. Benefiting from the epitaxy technique, the seamless stitching of these aligned domains leads to high crystalline quality of multilayer $MoS_2$ on sapphire, as will be confirmed by our latter device characterizations.

As mentioned above, $MoS_2$ multilayers would have N-dependent band gaps. In order to confirm it in our epitaxial samples, we thus collected optical spectra for our mono-, bi- and trilayer $MoS_2$ wafers. Corresponding Raman spectra are shown in Fig. 3a. In control samples of monolayer $MoS_2$ films, the peak frequency difference ($\Delta$) between the $E_{2g}$ and $A_{1g}$ vibration modes is about ~20 cm$^{-1}$. As a comparison, $\Delta$ in bilayer and trilayer films are wider to a number of ~23 and ~24 cm$^{-1}$, respectively. Fig. 3b shows the photoluminescence (PL) spectra of our mono-, bi- and trilayer $MoS_2$ films. We can see a strong A-exciton peak at ~1.88 eV in the monolayer, while A- and B-exciton peaks are greatly suppressed in bi- and trilayer films due to the transition from the direct band gap to the indirect ones[42,43]. The indirect bandgaps are of ~1.50 eV and ~1.42 eV for bilayers and trilayers, respectively, confirming the N-dependent band gaps of the multilayer $MoS_2$. Note that those sharp peaks at 1.79 eV are from sapphire substrates. Fig. 3c shows the optical transmittance spectra of mono-, bi-, and trilayer $MoS_2$ films transferred on quartz substrates and the corresponding transmittances are 94.2%, 91.6% and 84.5 % at a wavelength of ~550 nm. Due to the release of the strain after transfer, the A- and B-exciton peaks in the transmittance spectra are a little bit shifted. Using the Raman line scanning, we also investigated the wafer-scale uniformity of the as-grown mono-, bi- and trilayer $MoS_2$ wafers, as shown in Fig. 3 d-i. We can see these Raman peaks locate nearly the same along the entire wafer diameter, revealing a high uniformity.

Based on the obtained high-quality multilayer $MoS_2$ wafers, we hence fabricated FETs for performance benchmark testing. Let's look at the short channel trilayer $MoS_2$ FETs first. The structure of these back-gated $MoS_2$ FETs is illustrated in Fig. 4a. High-resolution STEM imaging at the $MoS_2$-Au interface (as illustrated in the bottom image of Fig. 4a) reveals a sharp contact interface without obvious damages, filamentous breaks or wrinkles[6,44-46]. The output and transfer curves of a device with channel length ($L_{ch}$) of 40 nm are shown in Fig. 4 b-c. Linear output characteristics at small bias voltages ($V_{ds}$) suggest the ohmic contact behavior, and the source-drain currents ($I_{ds}$) quickly approach to saturation at small gate voltages subjected to the employment of $HfO_2$ ($\varepsilon_r$=15-20) dielectric layer. The device features high on/off ratio of >10$^7$, sharp subthreshold swing (SS) of 200 mV/dec over 4 magnitudes, and small hysteresis of $\Delta V_g \approx 0.02$ V (at 0.1 μA·μm$^{-1}$). The current density ($I_{ds}$/W, where W is the channel width) can reach 1.70/1.22/0.94 mA·μm$^{-1}$ at $V_{ds}$=2/1/0.65 V which is the highest ever achieved in $MoS_2$ transistors.

Transfer curves of mono-, bi- and trilayer devices with $L_{ch}$=100 nm are shown in Fig. 4d. We can see a significant improvement of the on-current densities while increasing the number of layers, and the corresponding $I_{ds}$/W of mono-, bi- and trilayer devices are 0.40, 0.64 and 0.81 mA·μm$^{-1}$, respectively, at $V_{ds}$=1 V and $V_g$=5 V. It was also noted that thicker $MoS_2$ devices show saturated currents at much smaller $V_g$. In Fig. 4e, we plotted the current densities ($V_{ds}$=1 V) and on/off ratios of our devices, compared with previous data from the state-of-the-art $MoS_2$ devices. The good balance between high current density and high on/off ratio suggests a great potential of these epitaxial multilayer $MoS_2$ wafers for fabrication of integrated, high-performance and low-power electronics.

Next, we also fabricated long-channel FETs with $L_{ch}$ varying from 5 to 50 μm and $W_{ch}$ varying from 10 μm to 30 μm based on our multilayer $MoS_2$ wafers, as illustrated in the inset of Fig. 4f. Transfer curves of 150 randomly picked trilayer $MoS_2$ FETs with different $L_{ch}$ and $W_{ch}$ are shown in Fig. 4f. The overall yield of all devices is >95 %. All these devices exhibit small device-to-device variations, reflecting the uniformity of epitaxial wafers. On/off ratios, subthreshold voltages ($V_{th}$) and SS of these devices are also plotted in Fig. 4g. The highest on/off ratio can reach to $10^8$-$10^9$ and averages at $4.5 \times 10^8$, much higher than that achieved in the previous multilayer $MoS_2$ devices[32,35,47]. $V_{th}$ is mainly located at -1.25±0.4 V and the average SS is ~115 mV/dec.

Finally, let's compare film conductivities of mono-, bi- and trilayer $MoS_2$. The sheet resistances (ρ) were extracted by transfer length method (TLM)[48] as shown in Fig. 4h. At a carrier density of $n_i \approx 4 \times 10^{13}$ cm$^{-2}$, ρ is 9.3, 5.4 and 3.0 kΩ for mono-, bi- and trilayer $MoS_2$ channels, respectively, revealing that multilayer $MoS_2$ is more conductive. Besides, the extracted contact resistance ($R_c$) is ~0.61 kΩ·μm at $n_i \approx 4 \times 10^{13}$ cm$^{-2}$. Although the achieved $R_c$ is slightly larger than that of Bi-contacts reported recently[9], Au-contacts are advantageous considering that Au is stable and widely used in the nowadays semiconductor technology. Better device performances might be achievable in future by further optimizing contact techniques. In Fig. 4i, we summarize the field effect mobilities ($μ_{FE}$) of these long-channel $MoS_2$ FETs. A significant improvement on $μ_{FE}$ with channel layer numbers can be clearly seen, just as expected. The average $μ_{FE}$ is ~80, ~110 and ~145 cm$^2$·V$^{-1}$·s$^{-1}$ for mono-, bi- and trilayer FETs, respectively. The mobility distributions in each type of devices are fitted by Lorentz curves. The full width at the half maximum (FWHM) of the fitting is ~40, ~50 and ~60 cm$^2$·V$^{-1}$·s$^{-1}$ for mono-, bi- and trilayer devices, and the increased FWHM with number of layers is partially attributed to the inhomogeneity from additional layers and need to be optimized in further studies. Remarkably, the highest $μ_{FE}$ reaches 131.6, 217.3 and 234.7 cm$^2$·V$^{-1}$·s$^{-1}$ in our mono-, bi- and trilayer devices, and all these numbers are record-high in wafer scale $MoS_2$ devices. Considering that, in well-developed thin film transistors (TFTs), $μ_{FE}$ is 10-40 cm$^2$·V$^{-1}$·s$^{-1}$ for indium–gallium–zinc-oxide (IGZO) TFTs and 50-100 cm$^2$·V$^{-1}$·s$^{-1}$ for low-temperature polycrystalline silicon (LTPS) TFTs[49], the competitive average $μ_{FE}$ (larger than 100 cm$^2$·V$^{-1}$·s$^{-1}$) achieved in this work also reveal a great potential of these multilayer $MoS_2$ films for TFT applications.

As shown above, the developed layer-by-layer epitaxy on sapphire can yield uniform and large-scale multilayer $MoS_2$ with clean interfaces and well-controlled number of layers, e.g. 1, 2, 3. In each individual layer, the high lattice continuity/quality are accomplished via seamless stitching of large domains aligned along sapphire<11-20>. Bilayer and trilayer $MoS_2$ wafers exhibit remarkably improved electrical quality with respect to their monolayer counterparts, as evidenced from higher on-current densities and higher electron mobilities, suggesting a great potential of using them for 2D electronics. Regarding technological improvements, further investigations are required. Firstly, the high-temperature growth process is less compatible with the conventional semiconductor processes thus needs to be lowered. Secondly, steady improvements of wafer sizes and control of single-alignment of domains are also required for producing single-crystalline multilayers at large scale.

Besides, it is also very interesting to apply this layer-by-layer epitaxy technique for large-scale and high-quality heterogeneous 2D layers to broaden the application field of 2D semiconductors.

## Methods

*Layer-by-layer epitaxy of MoS$_2$.* All growths were carried out in a home-built multi-source CVD system with three temperature zones, named zone-I, zone-II and zone-III. In a typical growth, one S-source (Alfa Aesar, 99.9%, 15 g) was loaded in zone-I and carried by Ar (40 sccm) and six MoO$_3$-source (Alfa Aesar, 99.999%, 30 mg each) were loaded in zone-II and carried by Ar/O$_2$ (40/1.7 sccm) individually. Sapphire substrates (single side polished, c-plane (0001) with off-set angle (M-axis) of 0.2±0.1 deg., 4-inch wafers) were loaded in zone-III. During the heteroepitaxy of MoS$_2$ on sapphire, the temperature in zone-I, zone-II, and zone-III is kept at 120 ºC, 540 ºC and 910 ºC, respectively; while the temperature in zone-II and zone-III was increased to 570 ºC and 940 ºC, respectively, for homoepitaxy of MoS$_2$.

*Structural and spectroscopic characterizations.* AFM imaging was performed with Asylum Research Cypher S system. Raman and PL spectra were collected with Horiba Jobin Yvon LabRAM HR-Evolution Raman system with the excitation laser wavelength of 532 nm. SAED was performed in a TEM (JEOL Grand ARM 300 CFEG) operating at 80 kV, and atomic resolution images were achieved with an Aberration-corrected scanning transmission electron microscope Grand ARM 300 (JEOL) operating at 80 kV.

*Device Fabrications and Measurements.* FETs were fabricated by lithography and etching process. The device fabrication process is illustrated in Fig. S6. First, buried back-gates of Ti/Au/Ti (1/5/1 nm) were patterned on substrates by lithography and e-beam evaporation at a deposition rate of 0.01-0.05 Å/s. Second, HfO$_2$ with a thickness of 5-15 nm was deposited by ALD (Savannah-100 system, Cambridge NanoTech. Inc. Precursors: H$_2$O and tetrakis dimethylamino hafnium; Deposition temperature: 200 ºC) as the gate dielectric layer. Third, MoS$_2$ films were etched off from sapphire substrates in KOH solution (1 Mol/L) at 110 ºC and transferred onto the as-prepared HfO$_2$/metal-gate/sapphire surfaces. After transfer, lithography and oxygen plasma etching (Plasma Lab 80 Plus, Oxford Instruments Company) were used to define MoS$_2$ channel region. Finally, e-beam evaporated Au (20 nm) were deposited for source-drain contact metal. For short channel (L<100 nm) FETs, the substrate is SiO$_2$ and the channels were defined by standard e-beam lithography (EBL, Raith e-Line plus system) with PMMA (495 A2) as the resist layer (spin-coated at 2000-3000 rpm and baked at 180 ºC for 2 min). For long channel (L>2 μm) FETs, the substrate is sapphire and the channels were defined by UV-lithography (MA6, Karl Suss) with AR-P 5350 (ALLRESIST GmbH) as positive photoresist with thickness of ~1 μm (spin-coated at 4000 rpm and baked at 100 ºC for 4 min). Note that we also use oxygen plasma to clean the photoresist residues before depositing the Ti/Au/Ti back-gate electrodes before ALD. All electrical measurements were carried out in a four-probe vacuum station (base pressure ~10$^{-6}$ mbar) equipped with a semiconductor parameter analyzer (Agilent B1500).

# Acknowledgements

This work was supported by the National Key Research and Development Program, the Strategic Priority Research Program of CAS (Grant No. XDB30000000), the Key-Area Research and Development Program of Guangdong Province (Grant No. 2020B0101340001), the National Science Foundation of China (NSFC, Grant No. 11834017 & 61888102), and the Key Research Program of Frontier Sciences of CAS (Grant No. QYZDB-SSW-SLH004).

# Author Contributions

G.Z. supervised this research. Q.W. performed the CVD growths and Raman characterizations. J.T. carried out device fabrications and electrical measurements with the assistance from Q.W.. X.L., Q.Z., X.B., and L.G. performed STEM characterizations. J.L., and K.L. performed SHG mapping. D.J. and L.X performed modeling and theoretical calculations. Q.W., J.T. and G.Z. wrote and all authors commented on the manuscript.

## Conflict of Interest

The authors declare no conflict of interests.

## Data availability

The data that support the findings of this study are available from the corresponding authors on reasonable request.

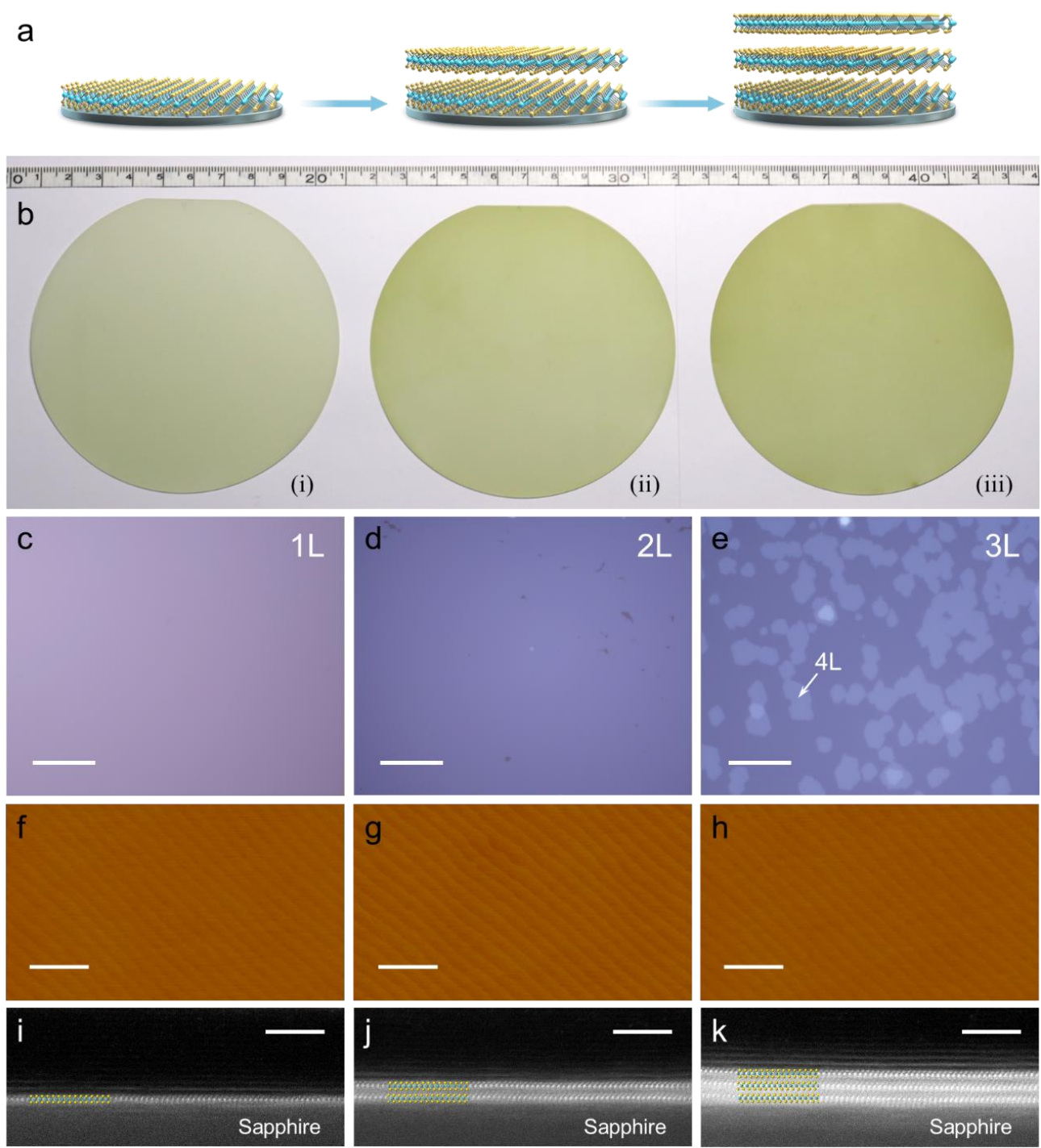

**Fig. 1 | Layer-by-layer epitaxy of multilayer MoS₂ wafers. a.** Schematic illustration of epitaxy process. **b.** Photographs of 4-inch MoS₂ wafers, (i) monolayer, (ii) bilayer, (iii) trilayer. **c-e.** Optical images of wafers shown in (**b**). Quadrilayer domains on the trilayer film is marked by white arrow. Scale bars, 30 μm. **f-h.** AFM amplitude images taken from mono-, bi- and trilayer wafers. Scale bars, 500 nm. **i-k.** Cross-sectional HAADF-STEM images of epitaxial mono-, bi- and trilayer MoS₂. Scale bars, 3 nm.

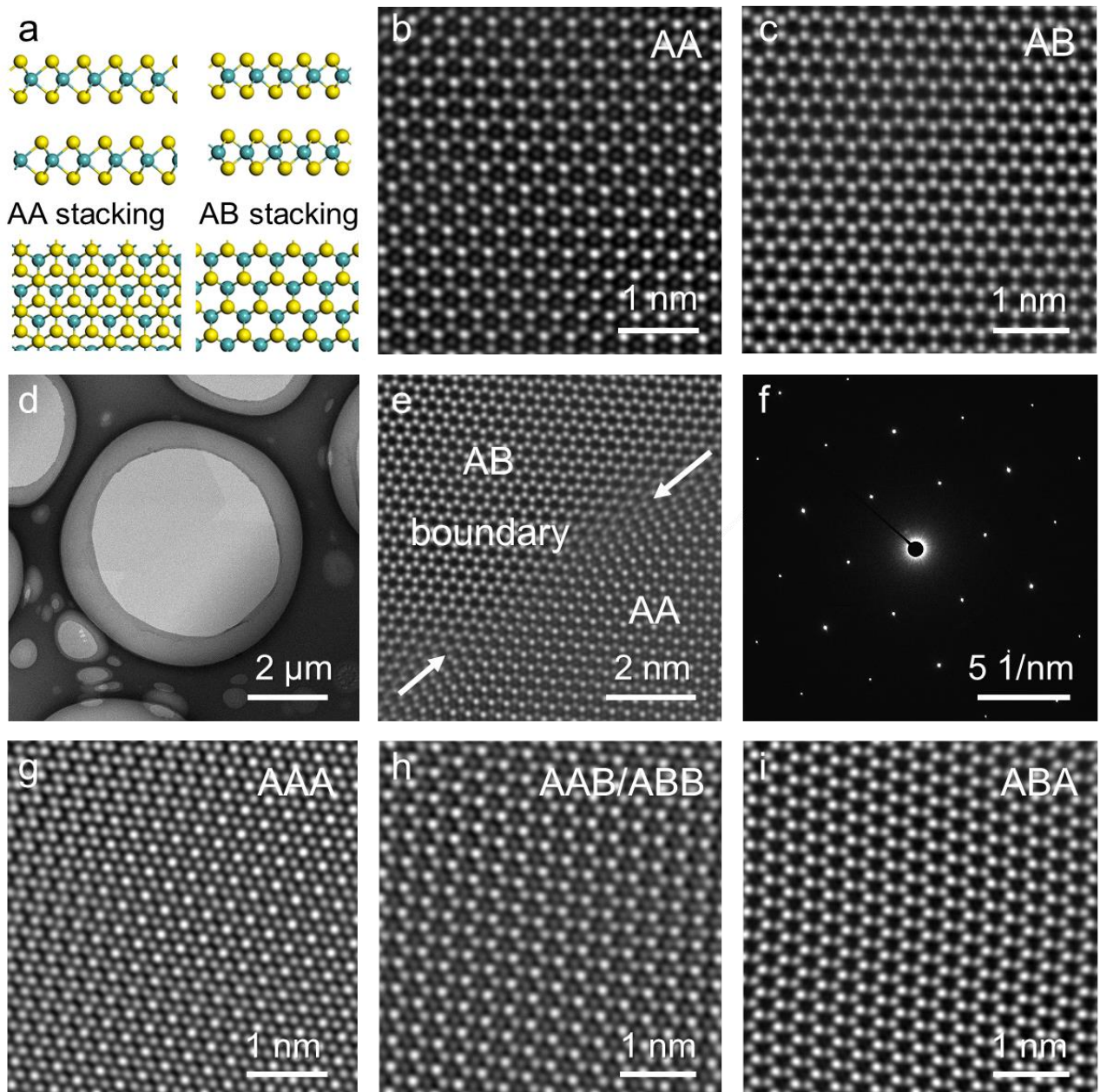

**Fig. 2 | Stacking configurations in the epitaxial multilayer MoS$_2$. a.** Side and top view in ball-and-stick mode of the atomic structures for AA and AB stacked MoS$_2$ bilayer. **b/c.** STEM images of AA/AB stacked bilayer MoS$_2$. **d.** STEM image of two emerged flakes with AA and AB stacking orders. **e/f.** STEM/SAED image of the boundary area shown in **d**. **g-i.** STEM images of the AAA stacked (**g**), AAB/ABB stacked (**h**) and ABA (**i**) stacked trilayer MoS$_2$.

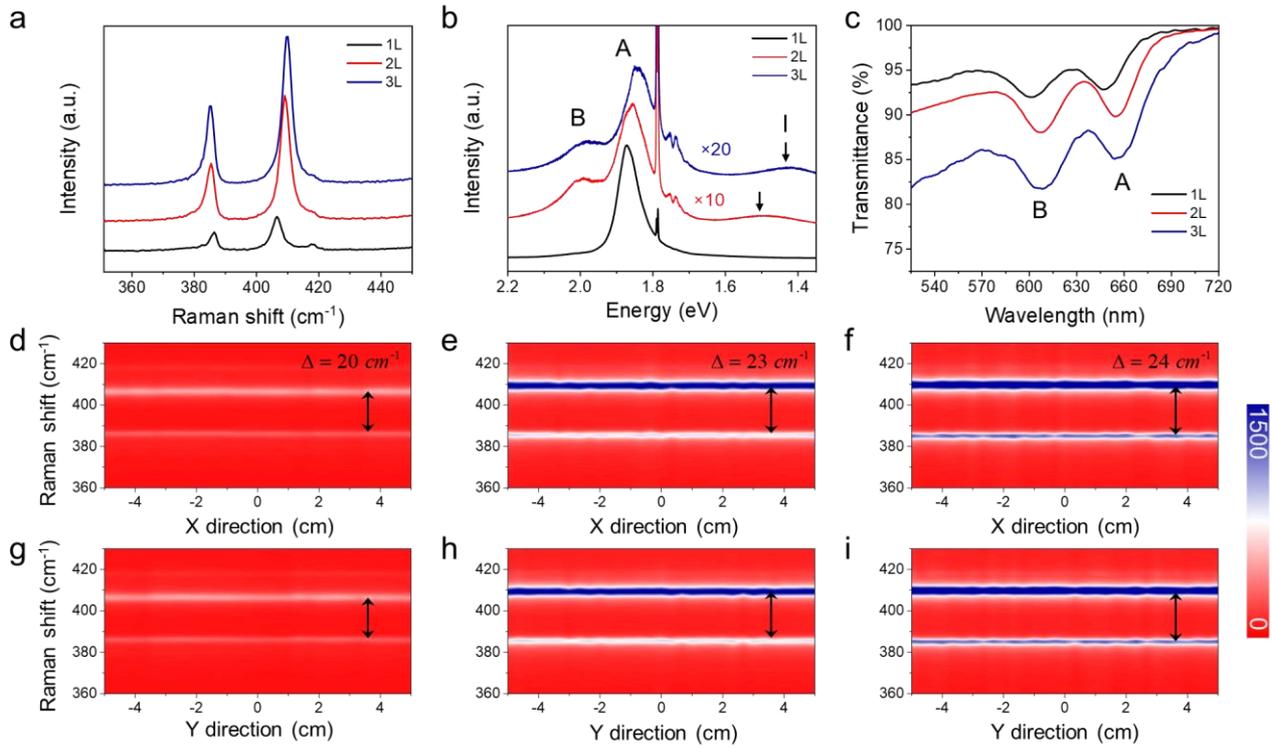

**Fig. 3 | Spatial uniformity of multilayer MoS$_2$ wafers. a-b.** Raman, PL, and transmittance spectra of the as-grown mono-, bi-, and trilayer MoS$_2$ wafers. **d-i.** Color-coded images of typical Raman line scan mapping along the horizontal and longitudinal direction of (**d**, **g**) monolayer, (**e**, **h**) bilayer and (**f**, **i**) trilayer MoS$_2$ wafers. Each line scan along either X- or Y-direction of the wafer includes 31 data points.

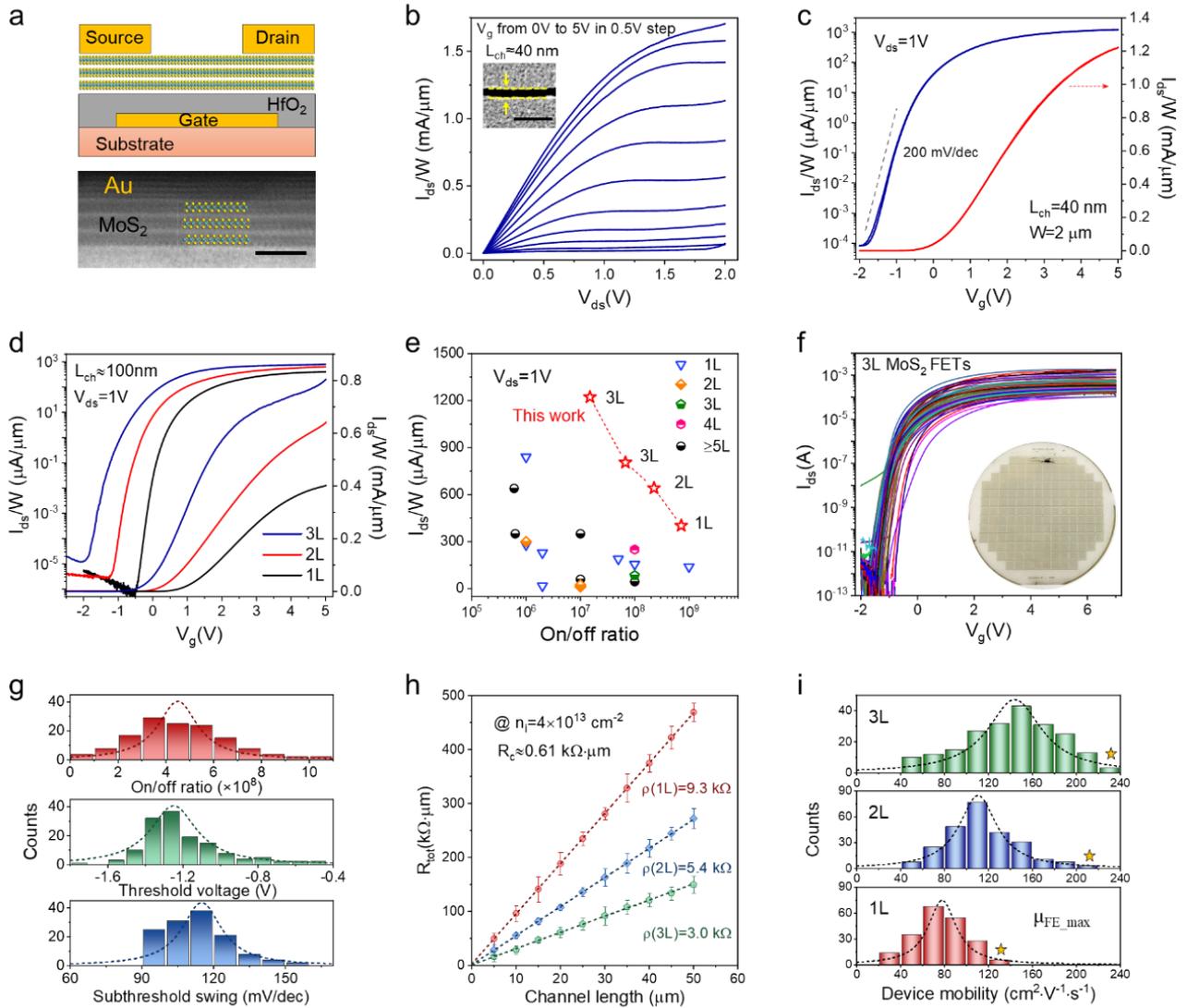

**Fig. 4 | Bench-mark testing of multilayer MoS$_2$ FETs. a.** Schematic view (top) of back-gated MoS$_2$ FET, and cross-section STEM image (bottom) of a trilayer FETs at the MoS$_2$-Au contact region. Scale bar, 1 nm. **b/c.** Typical output/transfer curves of a trilayer MoS$_2$ FET. L$_{ch}$=40 nm, t$_{HfO2}$=5 nm. Inset to (**b**) shows the SEM image of the channel. **d.** Comparison of transfer curves of mono-, bi- and trilayer MoS$_2$ FETs with L$_{ch}$≈100 nm. **e.** The comparisons of current densities (@V$_{ds}$=1 V) and on/off ratios with previous works. The detailed device parameters are shown in Table S1. **f.** Transfer curves of 150 trilayer MoS$_2$ FETs at V$_{ds}$=1 V. L$_{ch}$=5-50 μm, t$_{HfO2}$=10 nm. Inset to (**f**) shows photograph of wafer-scale MoS$_2$ FET array. **g.** Statistical distribution of on/off ratio (red), threshold voltage (green) and subthreshold swing (blue) from the 150 trilayer MoS$_2$ FETs. **h.** The sheet resistance ρ and contact resistance R$_c$ extracted from mono-, bi- and trilayer MoS$_2$ FETs. **i.** Statistical distribution of device mobility of mono-, bi- and trilayer MoS$_2$ FETs. The yellow stars indicate the maximum values achieved in each type of devices.